# Magneto-transport phenomena in p-doped diamond from first principles

Francesco Macheda[*] and Nicola Bonini[†]

*Department of Physics, King's College London, Strand, London WC2R 2LS, United Kingdom*

(Dated: August 31, 2018)

We present a first-principles study of the magneto-transport phenomena in p-doped diamond via the exact solution of the linearized Boltzmann transport equation, in which the materials' parameters, including electron-phonon and phonon-phonon interactions, are obtained from density functional theory. This approach gives results in very good agreement with experimental data for Hall and drift mobilities, low- and high-field magnetoresistance and Seebeck coefficient, including the phonon drag effect, in a range of temperatures and carrier concentrations. In particular, our results provide a detailed characterisation of the exceptionally high values for mobility and Seebeck coefficient, and predict a large magnetic-field driven enhancement of the Seebeck coefficient, of up to 30% in a magnetic field of 40 kOe already at room temperature.

Diamond is one of the most promising materials for high-performance and tough electronic devices for power electronics, sensors and high-energy-physics detectors [1]. This is due to its excellent thermo-mechanical, chemical and transport properties: it is hard, chemically inert and heat tolerant, and it has very high values of breakdown voltage, thermal conductivity and carrier mobility [2].

It is surprising that, despite an intense research effort in the field, there is still a large uncertainty about the precise upper bounds for the electrical transport coefficients of diamond and their dependence on doping and external fields (temperature and magnetic fields)—quantities of crucial importance to guide the optimisation of devices. For instance, the experimental values for the intrinsic hole drift mobility at room temperature vary significantly in literature, ranging from 2000 to 3800 cm$^2$V$^{-1}$s$^{-1}$ [3–8], and differ considerably from the reported Hall mobilities. In addition, the interplay between the charge carrier and phonon dynamics (both striking in diamond), that ultimately leads to an exceptionally large phonon drag effect [9], has not yet been precisely quantified.

In this work we determine the intrinsic transport properties of p-doped diamond, as well as their mutual dependence, and establish to what extent these properties can be controlled by doping, temperature and magnetic fields. For this we calculate transport coefficients by solving, for the first time, the Boltzmann transport equation (BTE) in presence of an arbitrary magnetic field, using ab initio materials' parameters as well as including the effect of phonon out-of-equilibrium populations (OEPs) that arise in presence of a temperature gradient.

The linearised BTE for electrons in presence of a small electric field **E**, a small temperature gradient $\nabla_\mathbf{r} T$ and an arbitrary magnetic field **B** is [10]:

$$-\frac{\partial f_k^0}{\partial \epsilon_k}\mathbf{v}_k \cdot \left[\mathbf{E}e + \frac{\nabla_\mathbf{r} T}{T}(\epsilon_k - \mu)\right] + D_k(g, \delta n)$$
$$+ \frac{e}{\hbar}\frac{\partial f_k^0}{\partial \epsilon_k}(\mathbf{v}_k \wedge \mathbf{B})\cdot \nabla_\mathbf{k}\chi_k = \frac{1}{N}\sum_{k'}\frac{\Pi^0_{k',k}}{k_B T}[\chi_{k'} - \chi_k] \quad (1)$$

where $N$ is the number of points in the Brillouin Zone (BZ), $\mu$ is the chemical potential, and $\mathbf{v}_k$, $\epsilon_k$ and $f_k^0$ are the velocity, the energy and the Fermi-Dirac distribution of an electron with quantum numbers $k$ (short for $m\mathbf{k}$, band index and wave vector); $D_k(g, \delta n)$ is the "phonon-drag" term, that arises from the coupling of the phonon OEPs ($\delta n$) with charge carriers (via the electron-phonon matrix $g$) and provides an extra driving force for the current (see Supplementary Material [11] and Ref. 18). On the r.h.s. of Eq. 1 there is the collision term, where $\Pi^0_{k',k}$ accounts for the scattering probabilities of the electrons with both lattice and impurities. The solution of the BTE gives the steady-state distribution of electrons $f_k = f_k^0 - \frac{\partial f_k^0}{\partial \epsilon_k}\chi_k$, from which the current density $\mathbf{J}^e$ can be computed together with the transport coefficients via $\mathbf{J}^e = \bar{\bar{\sigma}}(\mathbf{B})\left[\mathbf{E} + \bar{\bar{S}}(\mathbf{B})T\left(\frac{-\nabla_\mathbf{r} T}{T}\right)\right]$, where $\bar{\bar{\sigma}}(\mathbf{B})$ and $\bar{\bar{S}}(\mathbf{B})$ are the conductivity and the Seebeck tensors, respectively; in the linear regime for the magnetic field, the conductivity tensor can be expanded as $\sigma_{ij}(\mathbf{B}) = \sigma_{ij}^{(0)}(\mathbf{0}) + \sum_k \sigma_{ijk}^{(1)}(\mathbf{B})B_k$. In recent years there has been a significant effort to compute $\sigma_{ij}^{(0)}$ in bulk semiconductors from first principles both within the relaxation time approximation [12–15] as well as from the iterative solution of the BTE [16, 17]. In this work we solve Eq. 1 for $\mathbf{B} = 0$ using an efficient Conjugate Gradient (CG) solver, as discussed in Ref. 18. For $\mathbf{B} \neq 0$ we use an iterative approach to effectively deal with the term $\nabla_\mathbf{k}(\chi_k)$: we start from a guess for $\chi_k$, $\chi_k^{(0)}$, and then calculate the l.h.s. of Eq. 1; we can then solve Eq. 1 with the CG algorithm and obtain a new population, $\chi_k^{(1)}$; these operations are iterated until $|\sum_k \left(\chi_k^{(N+1)} - \chi_k^{(N)}\right)|^2$ is lower than a chosen threshold.

The electronic and vibrational properties of diamond were computed with density functional and density functional perturbation theory as implemented in QUANTUM ESPRESSO [19] within the local-density approximation (LDA) [20]. We used a norm-conserving pseudopotential and a plane-wave expansion up to a 60 Ry, a BZ sampling with a 24 × 24 × 24 Monkhorst-Pack mesh and a theoretical lattice parameter of 3.52 Å. Electronic bands and electron-phonon matrix elements have been first calculated on grids of $10 \times 10 \times 10$ k-points for electrons and $5 \times 5 \times 5$ q-points for phonons. Then we have used the Wannier interpolation scheme as implemented in the WANNIER90 [21] and EPW [22, 23] codes to compute transport quantities on very dense meshes up to $100 \times 100 \times 100$. The phonon OEPs were obtained with the D3Q [24] and THERMAL2 [25] codes. We used the model for boron doping of Ref. 26, and scattering from ionized impurities is dealt with the Brooks and Herring formula [27], while the one from neutral impurities is treated with the model in Ref. 28.

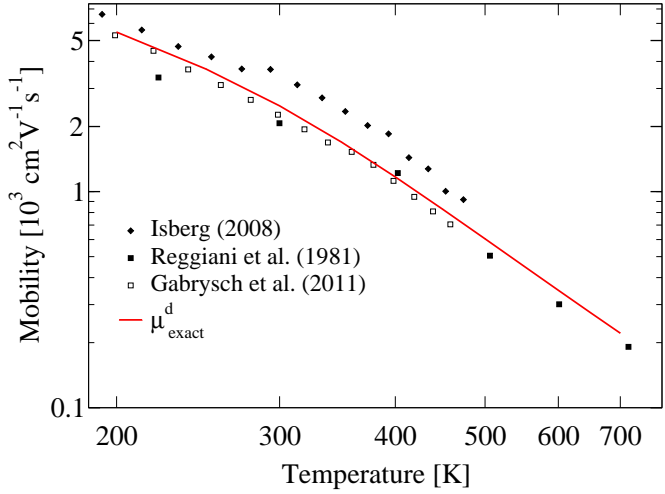
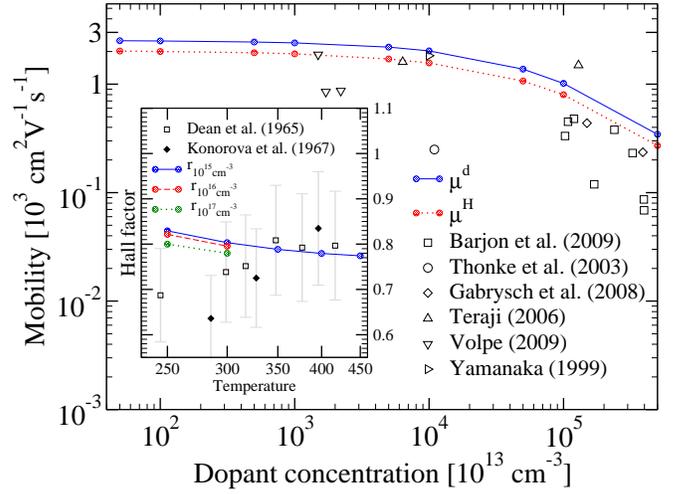

FIG. 1. Temperature dependence of hole drift mobility in p-doped diamond with a boron concentration of $10^{15}$ cm$^{-3}$ (red solid line). Experimental data from Ref. [8] (black squares), [29] (white squares), [3] (diamonds).

FIG. 2. Doping dependence of $\mu^d$ and $\mu^H$ in p-doped diamond at 300 K. Experimental data for Hall mobility taken from Ref. [35] (circles), [36] (left triangles), [37] (up triangles), [30] (diamonds), [38] (down triangles), [39] (squares). Insert: Hall coefficient factor as function of temperature for Boron concentrations of $10^{15}$ cm$^{-3}$ (blue circles), $10^{16}$ cm$^{-3}$ (light blue circles) and $10^{17}$ cm$^{-3}$ (green circles). Experimental data taken from Ref. [40](squares) and [32](diamonds): the data were obtained by comparing $\mu^d$ and $\mu^H$ of similar samples and an uncertainty of around $\pm 15\%$ was estimated. [41]

Fig. 1 shows the low doping (boron density of $10^{15}$ cm$^{-3}$) hole drift mobility of diamond, $\mu^d = \sigma_{xx}/(en)$ (where $n$ is the carrier density), between 200 K and 700 K. Our calculations predict a room-temperature mobility of around 2500 cm$^2$ V$^{-1}$ s$^{-1}$. This is in good agreement with the most recent measurements reported in literature by Gabrysch et al. on CVD diamond with very low dopant concentration [29]. Our result confirms that p-doped diamond has a very high hole mobility, but it also indicates that the extraordinary values (between 2600 and 3800 cm$^{-2}$ V$^{-1}$ s$^{-1}$) reported by Isberg and coworkers [4, 30] are not compatible with our ab initio predictions. This is especially true if we consider that our ab initio results can be regarded as an upper bound to the mobility achievable in p-doped diamond. Indeed, at such low doping the result is very close to the intrinsic phonon-limited value, and, since LDA tends to underestimate the electron-phonon coupling in diamond [31] and the mobility in other semiconductors [17], our LDA results could slightly overestimate the exact value for the mobility.

The temperature dependence of the mobility shows a change in slope between 250 K and 400 K: at high temperatures, the theoretical drift mobility is proportional to $\sim T^{-2.95}$ whereas at low temperatures it is proportional to $\sim T^{-1.70}$. This is in good agreement with the results by Gabrysch et al. [29] for CVD samples and consistent with values obtained by Reggiani et al. [32] for natural diamond. The analysis of the different scattering mechanisms (see Supplementary Material) shows that the low temperature behaviour is determined by the acoustic phonon scattering; here the ab initio description of bands and phonon scattering results in a slope of the mobility that is slightly steeper than the typical $T^{-\frac{3}{2}}$ deduced using models based on spherical energy bands and elastic scattering processes. [33] The change in slope observed at high temperatures is instead due to the onset of optical phonon scattering, which in diamond is exceptionally strong. For instance, we find that the zone centre optical modes contribute to about 14% of the resistivity at 300 K, while at around 500 K their contribution becomes similar to the acoustic modes. For comparison, in n-doped silicon at 300 K the zone centre optical modes give almost no contribution to the resistivity ([18]) despite the fact that their frequency is $\sim 2.5$ times smaller than that of the optical modes in diamond. In passing, it is worth noticing that in spite of the faster decrease of the mobility at high temperature, diamond still exhibits mobilities of around 500 cm$^2$V$^{-1}$s$^{-1}$ at 500 K. These are very high values especially when compared to other wide bandgap semiconductors important for high-temperature electronics (for instance 4H-SiC has a hole mobility of around 30 cm$^2$V$^{-1}$s$^{-1}$ at 500 K [34]).

To complete the picture of the mobility, we analyse this quantity as a function of boron concentration. As in doped samples the Hall mobility, $\mu^H = \mu^d ne\sigma_{xyz}^{(1)}/(\sigma_{xx}^{(0)})^2$, rather than the drift mobility, is usually measured, in Fig. 2 we present both $\mu^d$ and $\mu^H$ for acceptor concentrations ranging from $10^{15}$ cm$^{-3}$ to $5\cdot10^{18}$ cm$^{-3}$ at 300 K. Our results show that for carrier concentrations lower than $10^{15}$ cm$^{-3}$, $\mu^d$ and $\mu^H$ are mainly limited by electron-phonon interactions, while at higher carriers concentrations they decrease as a result of an increased electron-impurity scattering. Here the comparison with experiments is not straightforward. Indeed, the reported experimental mobilities can be spread out over one order of magnitude at a given B-content. This is a consequence of compensation effects due to deep level impurities. These effects can greatly affect electrical transport as they tend to both decrease the carrier concentration and increase the density of ionised impurities that can scatter charge carriers. However, Fig. 2 shows that our mobilities lie above almost all the measured data in a wide range of doping concentrations. This is not surprising as in our calculation we have neglected compensation effects and since, as mentioned above, our results should be regarded as an upper limit to the achievable mobility. Also, it is clear that $\mu^H$ is significantly lower than $\mu^d$ and closer to the higher experimental Hall mobilities reported in this range of concentrations. The difference between $\mu^d$ and $\mu^H$ is analysed in more detail in the insert of Fig. 2: the Hall factor, $\mu^H/\mu^d$, is around $\sim 0.8$ and it is fairly insensitive to doping and temperature. The possible decrease of the Hall factor at low temperatures observed in experiments might be due to electron-impurity scattering, which is the mechanism that affects the most the Hall factor in this temperature range. Overall, our results validate the analysis of the experimental data on diamond in Ref. 41.

As our theoretical approach provides access to transport properties at arbitrary magnetic fields, a more detailed comparison between theory and experiment can be done by focussing on the magnetoresistance, $[\rho_{ij}(B_l) - \rho_{ij}(0)]/\rho_{ij}(0)$, where $\bar{\bar{\rho}}(\mathbf{B}) = \bar{\bar{\sigma}}(\mathbf{B})^{-1}$ is the resistivity tensor. Fig. 3 shows the values of the transversal (TM) and longitudinal magnetoresistance (LM) as a function of $|\mathbf{B}|$. We find that the LM is always smaller than the transversal one, and that both are proportional to $|\mathbf{B}|^2$ at low fields (as expected from symmetry considerations [42]), while at higher fields the $|\mathbf{B}|$ dependence follows a lower slope. The magnitude of the magnetoresistance is quite similar to what found experimentally in p-type Si and Ge (the only exception is the TM of p-type Ge which is about one order of magnitude larger than in diamond) [43]. The results are in overall good agreement with the experimental data, and the differences between theory and experiment might be related to the unclear doping content of the measured samples and the fact that the experimental data have been taken with respect to different orientations of $\mathbf{J}^e$ and $\mathbf{B}$. To give an idea of the effective strength of the magnetic fields considered here, in the insert of Fig. 3 (and of Fig. 5 below) we show the values for the Hall angle, i.e. the angle between $\mathbf{E}$ and $\mathbf{J}^e$, as a function of the magnetic field strength. This quantity is expected to reach $\frac{\pi}{2}$ at very large magnetic fields when the magnetoresistivity saturates [46].

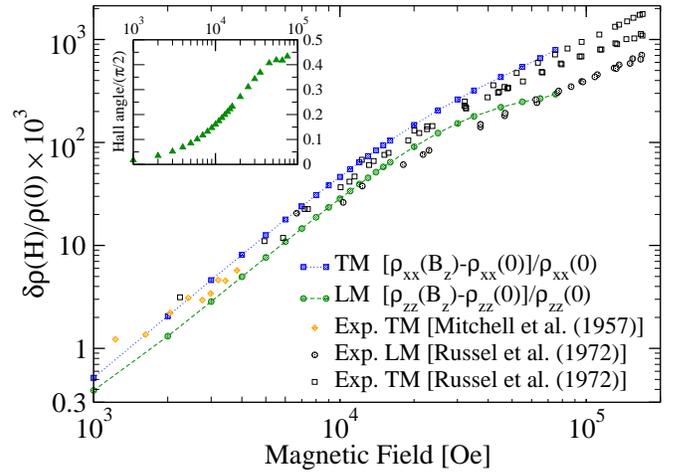

FIG. 3. Transversal magnetoresistance (TM) and longitudinal magnetoresistance (LM) as a function of the magnetic field strength at 300 K (boron concentration of $10^{15}$ cm$^{-3}$). Experimental data are from Ref. [44] (orange squares) and Ref. [45] (black squares and circles). Insert: Hall angle as a function of the magnetic field.

Controlling the mobility with $\mathbf{B}$ provides a way to enhance the thermoelectric transport properties. To see this, let's focus on the effect of an applied thermal gradient and in particular on the Seebeck coefficient, $S = S_{xx}$. Fig. 4 shows that the phonon drag contribution to the Seebeck coefficient, $S_p$, due to $D_k$ in the BTE, leads to extremely high values of $S$ at low temperature: for instance, at 200 K $S$ is five times larger than the value reported for silicon [18]. We notice that even at 700 K the phonon drag represents a sizeable contribution to the already large diffusive part of $S$. Such high Seebeck coefficient might be of potential interest for temperature sensing applications, especially for devices operating in harsh environments. Overall, our results are in good agreement with the available experimental data. The difference from the measured values, more evident at low temperatures, could be due to the interpolation method used for the phonon OEPs, as well as to possible inaccuracies related to the use of the LDA functional. The analysis of the phonon drag contribution (see Supplementary Material) shows that up to 700 K this quantity is mostly determined by acoustic phonons with frequencies below 300 cm$^{-1}$. A decrease in the temperature reduces the subset of acoustic modes responsible for the phonon drag: for instance, at 200 K there is no contribution from phonons above 100 cm$^{-1}$. It is worth pointing out that $S$ is practically unaffected by inclusion of isotopic scattering in the calculation of the phonon OEPs. The reason is that isotopic scattering affects almost only the OEP of high-frequency phonons (that leads to a substantial decrease of the lattice thermal conductivity, especially at low temperatures), but these have no significant impact on $S$.

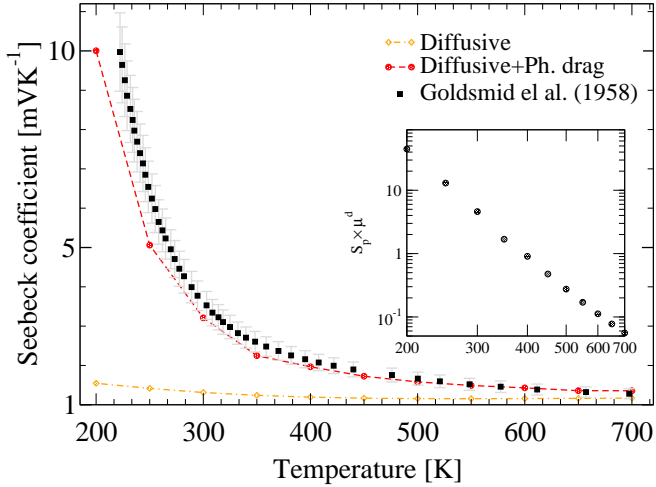 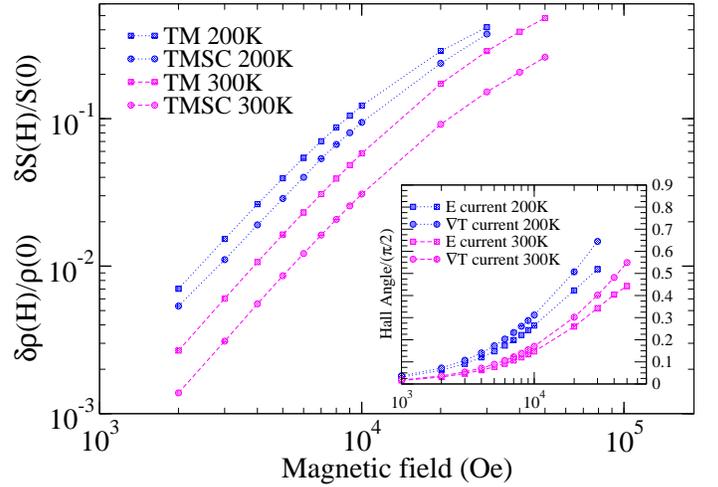

FIG. 4. Temperature dependence of the total Seebeck coefficient (diffusive+phonon-drag) (red circles) and of the diffusive contribution (orange diamond) in p-doped diamond (boron concentration of $10^{15}$ cm$^{-3}$). The experimental data (including a ±10% error bar) are from [9]. Insert: phonon drag component ($S_p$) times mobility as function of temperature.

FIG. 5. Transversal magnetoresistance (TM) and transversal magneto-Seebeck coefficient (TMSC) as a function of the magnetic field strength, at 200 K and 300 K (boron concentration of $10^{15}$ cm$^{-3}$). Insert: Hall angle for the electric and thermal gradient contributions to the total electronic current as a function of the magnetic field.

It is interesting to note that, when the the phonon drag term $D_k$ is the dominant driving force in Eq. 1, the Seebeck coefficient is expected to be inversely proportional to the mobility. This can be seen from the insert of Fig. 4 where we show that the product of $S_p$ and $\mu^d$ is proportional to $\sim T^{-5.5}$ (the exponent is slightly higher at high temperatures). This result is in good agreement with the dependence obtained by Herring for semiconductors using simplified models for bands and scattering relaxation times [47]. The interrelation between $S_p$ and $\mu^d$ can be exploited to further increase the Seebeck coefficient of diamond. This can be achieved when a temperature gradient is combined with a magnetic field that, as shown previously, tends to increase the resistivity. In order to quantify the extent of this effect in diamond we computed the transversal magneto-seebeck coefficient, $[S_{xx}(B_z) - S_{xx}(0)]/S_{xx}(0)$. This quantity and the TM are shown in Fig. 5. As expected, the magnetic field dependence of the two quantities is very similar, especially at low fields. While a similar trend has been observed in other semiconductors, such as germanium [48, 49], it's interesting to inspect here the magnitude of the enhancement of the Seebeck coefficient due to the applied magnetic field. For instance, we predict a relative change in $S$ at 200 K and 10 kOe that is about 4 times larger than in n-type germanium at the same temperature and magnetic field, even though the TM in n-type Ge is around 3 times bigger [48]. More interestingly, the effect is significant also at relatively high temperatures: indeed, the enhancement goes up to about 30% in a magnetic field of 40 kOe already at 300 K.

In conclusion, we have explored from first-principles the inherent limits of the thermo-electric properties of p-doped diamond, including their dependence on doping, temperature and magnetic fields. Our results provide a detailed microscopic characterisation of the dynamics of carriers in this material and represent a useful reference for the analysis and simulation of diamond-based electronic devices. Finally, we expect that the novel approach used and validated in this work will provide a valuable tool to address magneto-transport phenomena both in doped semiconductors as well as in metallic systems.

We acknowledge the ARCHER UK National Supercomputing Service and the Cirrus UK National Tier-2 HPC Service at EPCC.

* francesco.macheda@kcl.ac.uk
† nicola.bonini@kcl.ac.uk
[1] R. S. Balmer et al., J. Phys.-Condens. Matt. **21**, 364221 (2009)
[2] C. J. Wort et al., Mater. Today **11**, 22 (2008)
[3] J. Isberg et al., Science **297**, 1670 (2002)
[4] J. Isberg et al., Phys. Status Solidi A **202**, 2194 (2005)
[5] H. Pernegger et al., J. Appl. Phys. **97**, 073704 (2005)

[6] M. Nesladek et al., Diam. Relat. Mater. **17**, 1235 (2008)
[7] H. Jansen et al., J. Appl. Phys. **113**, 173706 (2013)
[8] L. Reggiani, S. Bosi, C. Canali, F. Nava and S.F. Kozlov, Phys. Rev. B **23**, 3050 (1981)
[9] H. J. Goldsmid, P. Phys. Soc. **73**, 393 (1959)
[10] J. M. Ziman, *Electrons and Phonons: The Theory of Transport Phenomena in Solids*, (Oxford university Press, London, 2001)
[11] See Supplementary Material at [URL]
[12] F. Murphy-Armando and S. Fahy, Phys. Rev. B **78**, 035202 (2008)
[13] O. D. Restrepo et al., Appl. Phys. Lett. **94**, 212103 (2009)
[14] J. Zhou et al., Proc. Natl. Acad. Sci. USA **112**, 14777 (2015)
[15] J.J. Zhou and M. Bernardi, Phys. Rev. B **94**, 201201(R) (2016)
[16] J. Ma, A. S. Nissimagoudar and W. Li, Phys. Rev. B **97**, 045201 (2018)
[17] S. Poncé, E. R. Margine and F. Giustino, Phys. Rev. B **97**, 121201(R) (2018)
[18] M. Fiorentini and Nicola Bonini, Phys. Rev. B B **94**, 085204 (2016)
[19] P. Giannozzi et al., J. Phys.-Condens. Matt. **21**, 395502 (2009)
[20] J.P. Perdew and A. Zunger, Phys. Rev. B **23**, 5048 (1981)
[21] A. A. Mostofi et al., Comput. Phys. Commun. **178**, 685 (2008)
[22] F. Giustino, M.L. Cohen and S.G. Louie, Phys. Rev. B **76**, 165108 (2007)
[23] S. Poncé, E. R. Margine, C. Verdi and F. Giustino, Comput. Phys. Commun. **209** (2016)
[24] L. Paulatto, F. Mauri and M. Lazzeri, Phys. Rev. B **87**, 214303 (2013)
[25] G. Fugallo, M. Lazzeri, L. Paulatto and F. Mauri, Phys. Rev. B **88**, 045430 (2013)
[26] F. Fontaine, J. Appl. Phys. **85**, 1409 (1999)
[27] C. Jacoboni et al., Rev. Mod. Phys. **55**, 645 (1983)
[28] J.R. Meyer and F.J. Bartoli, Phys. Rev. B **24**, 2089 (1981)
[29] M. Gabrysch et al., J. Appl. Phys. **109**, 063719 (2011)
[30] M. Gabrysch et al, Phys. Status Solidi A **205**, 2190 (2008)
[31] G. Antonius, S. Ponce, P. Boulanger, M. Cote and X. Gonze, Phys. Rev. Lett. **112**, 215501 (2014)
[32] P. J. Dean et al., Phys. Rev. **140**, A352 (1965)
[33] L. Gherardi et al., Lett. Nuovo Cimento (1971-1985) **14**, 225 (1975)
[34] A. Koizumi et al., J. Appl. Phys. **106**, 013716 (2009)
[35] K. Thonke, Semicond. Sci. Tech. **18**, S20 (2003)
[36] S. Yamanaka et al., Phys. Status Solidi A **174**, 59 (1999)
[37] T. Teraji et al., Diam. Rela. Mater. **15**, 602 (2006)
[38] P.-N. Volpe et al., Appl. Phys. Lett. **94**, 092102 (2009)
[39] J. Barjon et al., Phys. Status Solidi A **3**, 202 (2009)
[40] E. Konorova et al., Fiz. Tekh. Poluprovodn. (Sov. Phys.-Semicond.) **1** (1967)
[41] L. Reggiani, D. Waechter and S. Zukotynski, Phys. Rev. B **28**, 3550 (1983)
[42] W. Jones and N. H. March, *Theoretical solid state physics* (Wiley, London, 1973)
[43] H. Nakagawa et al., J. Appl. Phys. **56**, 6 (1984)
[44] E. W. J. Mitchell et al., P. Phys. Soc. **70**, 527 (1957)
[45] K. J. Russel et al., Phys. Rev. B **6**, 4588 (1972)
[46] J. A. Swanson, Phys. Rev. **99**, 1799 (1955)
[47] C. Herring, Phys. Rev. **96**, 1163 (1954)
[48] M. C. Steele, Phys. Rev. **107**, 81 (1957)
[49] C. Herring et al., Phys. Rev. **111**, 36 (1958)